\documentclass[12pt,english]{article}
\pdfoutput=1
\usepackage[T1]{fontenc}
\usepackage{amssymb}
\usepackage{amsmath}
\usepackage{cite}
\usepackage{appendix}
\usepackage{epsfig}
\usepackage[unicode=true,
 bookmarks=true,bookmarksnumbered=false,bookmarksopen=false,
 breaklinks=false,pdfborder={0 0 1},backref=false,colorlinks=true]
 {hyperref}
\hypersetup{pdftitle={Warping Wormholes with Dust: a Metric Construction of the Python's Lunch},
 pdfauthor={Ning Bao, Aidan Chatwin-Davies, Grant N. Remmen},
 citecolor=black,linkcolor=black,urlcolor=black}
\usepackage{breakurl}
\usepackage[hang,flushmargin]{footmisc}
\usepackage{breakurl}
\usepackage{color}

\makeatletter
\def\simgt{\mathrel{\lower2.5pt\vbox{\lineskip=0pt\baselineskip=0pt
           \hbox{$>$}\hbox{$\sim$}}}}
\def\simlt{\mathrel{\lower2.5pt\vbox{\lineskip=0pt\baselineskip=0pt
           \hbox{$<$}\hbox{$\sim$}}}}
\makeatother

\newcommand{\be}{\begin{equation}}
\newcommand{\ee}{\end{equation}}
\newcommand{\bea}{\begin{eqnarray}}
\newcommand{\eea}{\end{eqnarray}}

\newcommand{\Ref}[1]{Ref.~\cite{#1}}
\newcommand{\Fig}[1]{Fig.~\ref{#1}}
\newcommand{\Eq}[1]{Eq.~\eqref{#1}}
\newcommand{\Eqs}[2]{Eqs.~\eqref{#1} and \eqref{#2}}
\newcommand{\Sec}[1]{Sec.~\ref{#1}}
\newcommand{\Secs}[2]{Secs.~\ref{#1} and \ref{#2}}

\DeclareMathOperator{\sech}{sech}
\DeclareMathOperator{\arccosh}{arccosh}
\DeclareMathOperator{\arcsinh}{arcsinh}

\newcommand{\mrm}[1]{\mathrm{#1}}

\newcommand{\dee}{{\rm d}}

\newcommand{\email}[1]{\href{mailto:#1}{\nolinkurl{#1}}}

\definecolor{darkgreen}{rgb}{0.0, 0.5, 0.0}

\makeatletter

\def\lambdabar{\protect\@lambdabar}
\def\@lambdabar{%
\relax
\bgroup
\def\@tempa{\hbox{\raise.73\ht0
\hbox to0pt{\kern.25\wd0\vrule width.5\wd0
height.1pt depth.1pt\hss}\box0}}%
\mathchoice{\setbox0\hbox{$\displaystyle\lambda$}\@tempa}%
{\setbox0\hbox{$\textstyle\lambda$}\@tempa}%
{\setbox0\hbox{$\scriptstyle\lambda$}\@tempa}%
{\setbox0\hbox{$\scriptscriptstyle\lambda$}\@tempa}%
\egroup
}

\makeatother

\begin{document}

\interfootnotelinepenalty=10000
\baselineskip=18pt

\hfill

\vspace{2cm}
\thispagestyle{empty}
\begin{center}
{\LARGE \bf
Warping Wormholes with Dust:
}\\
{\large a Metric Construction of the Python's Lunch}\\
\bigskip\vspace{1cm}{
{\large Ning Bao,${}^{a,b}$ Aidan Chatwin-Davies,$^{c}$ and Grant N. Remmen${}^{b}$}
} \\[7mm]
 {\it ${}^a$Computational Science Initiative\\[-1mm] Brookhaven National Laboratory, Upton, New York, 11973 \\[1.5 mm]
 ${}^b$Center for Theoretical Physics and Department of Physics \\[-1mm]
     University of California, Berkeley, CA 94720 and \\[-1mm]
     Lawrence Berkeley National Laboratory, Berkeley, CA 94720 \\[1.5mm]
     ${}^c$KU Leuven, Institute for Theoretical Physics\\[-1mm]
 Celestijnenlaan 200D B-3001 Leuven, Belgium} \let\thefootnote\relax\footnote{\noindent e-mail: \email{ningbao75@gmail.com}, \email{aidan.chatwindavies@kuleuven.be}, \email{grant.remmen@berkeley.edu}} \\
 \end{center}
\bigskip
\centerline{\large\bf Abstract}
\begin{quote} \small
We show how wormholes in three spacetime dimensions can be customizably warped using pressureless matter.
In particular, we exhibit a large new class of solutions in $(2+1)$-dimensional general relativity with energy-momentum tensor describing a negative cosmological constant and positive-energy dust.
From this class of solutions, we construct wormhole geometries and study their geometric and holographic properties, including Ryu-Takayanagi surfaces, entanglement wedge cross sections, mutual information, and outer entropy.
Finally, we construct a Python's Lunch geometry: a wormhole in asymptotically anti-de~Sitter space with a local maximum in size near its middle.
\end{quote}

\setcounter{footnote}{0}

\newpage
\tableofcontents
\newpage

\section{Introduction}

Much progress has been made over the past fifteen years in the study of the intricate relationship between quantum information theory and quantum gravity, especially within the context of the Anti-de~Sitter/Conformal Field Theory (AdS/CFT) correspondence \cite{Maldacena:1997re,Witten:1998qj,Aharony:1999ti}.
Holographic dualities between information theoretic quantities and bulk geometric quantities form a central part of this program, the most prominent perhaps being the duality between extremal surface area and entanglement entropy \cite{Ryu_2006,Hubeny:2007xt}.
Proposals for the holographic duals of myriad other entropic quantities have since followed, as well as conjectures about the holographic dual of entanglement itself~\cite{Maldacena_2013} and other information theoretic quantities, such as computational complexity~\cite{Brown_2016, Brown2_2016,susskind2014addendum}.

\begin{figure}[t]
    \centering
 \includegraphics[width=0.5\textwidth]{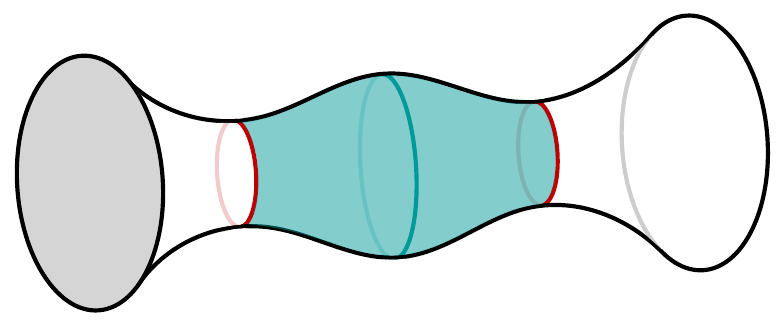}
    \caption{Schematic sketch of a Python's Lunch wormhole geometry. The wormhole's width exhibits two local minima (red), between which is a region (teal shading)  where reconstruction using boundary data on either side of the wormhole alone is obstructed. This is because once a local minimum exists in the bulk geometry homologous to an entire boundary, no minimal surface anchored to that boundary can probe past the local minimum, thus precluding any sort of bulk reconstruction that relies on boundary-anchored extremal objects, such as in Ref.~\cite{BCFK_2019}. Another way of seeing this difficulty in the context of operator reconstruction is through the greedy algorithm reconstruction of Ref.~\cite{Pastawski_2015}; attempting to push an operator in the bulging region outward results in the algorithm becoming trapped at the waists.}
    \label{fig:PythonLunchFig}
\end{figure}

Exact gravitational solutions provide rich bottom-up tests and insight into holographic proposals and conjectures away from the AdS vacuum.
In this work, we present a new class of asymptotically AdS spacetimes in $2+1$ dimensions that contain pressureless dust, with a spatial distribution that is straightforwardly customizable; see \Eqs{eq:metric}{eq:rho}.
Moreover, for suitable choices of the distribution, one can fold up slices of these spacetimes into two-sided wormholes that have irregular profiles in their throats.
Among this new class of spacetimes, we will find that the dust distribution can be engineered so as to generate a concrete implementation of a ``Python's Lunch,'' a special type of wormhole proposed in \Ref{Brown:2019rox} and sketched in \Fig{fig:PythonLunchFig}, which will allow our classical construction to make contact with recent investigations into holographic complexity.

The question of quantum mechanical complexity associated with gravitational systems has been of interest for some time in the context of black holes and holography.
In Ref.~\cite{Harlow_2013}, Harlow and Hayden argued that certain tasks, in particular the distillation of a Bell pair's worth of information from Hawking radiation, is exponentially hard in the evaporating black hole's entropy.
However, the subsequent ``complexity equals action'' conjecture of Refs.~\cite{Brown_2016, Brown2_2016,susskind2014addendum} generated tension with this argument.
According to the conjecture, given a holographic CFT state that is dual to a two-sided black hole, such as the thermofield double, its computational complexity is dual to the length of the wormhole or to the value of the gravitational action within the Wheeler-DeWitt patch.
However, the Harlow-Hayden proposal can be reformulated as the protocol necessary to prepare the thermofield double state by acting on the two CFTs in a separable manner, where one of the CFTs plays the role of the black hole and the other the emitted Hawking radiation.
On these grounds, one therefore concludes that the complexity of the thermofield double state is very large, while the complexity equals action proposal yields a much smaller value given by the gravitational action.

The Python's Lunch proposal of Ref.~\cite{Brown:2019rox} suggested that the source of this discrepancy could be the restriction that the computation described by \Ref{Harlow_2013} only makes use of computational gates that act on the two CFTs (or equivalently, the two sides of the wormhole) in a separable manner, i.e., not in a coupled fashion.
In other words, the complexity that appears in the Harlow-Hayden protocol is a restricted complexity, which could clearly be much larger than an ``unrestricted'' complexity measured by gravitational action.
The Python's Lunch was then offered as an example to support this resolution with starker geometric intuition.

In essence, a Python's Lunch geometry is an asymptotically AdS wormhole with a bulge in its throat.
A spacelike slice of such a wormhole has a local maximum in its width near the middle of the throat, with local minima on either side, as illustrated in \Fig{fig:PythonLunchFig}.\footnote{A related concept is Wheeler's ``bag of gold'' spacetime; see \Ref{Marolf:2008tx} and refs. therein. While the bag of gold requires a high degree of symmetry to stitch the FRW cosmology onto a Schwarzshild black hole throat, the Python's Lunch construction we will explore will not require this constraint.}
The region in between the local minima is difficult to access from one boundary alone, in the sense that once such a narrow waist exists that is homologous to an entire boundary, minimal boundary-anchored objects do not probe past the waist.
Provided that small numbers of individual computational operations on a single boundary correspond to bulk operations with negligible backreaction in the region swept out by boundary-anchored probes, it would therefore take a large number of such operations to probe the inner region.
In other words, the probes for which it is easy to access the inner region are those probes that are anchored to both boundaries.

This proposal is compelling, and so it is desirable to explicitly construct Python's Lunch geometries that go beyond the special-case constructions and tensor network analogies discussed  in \Ref{Brown:2019rox}. Moreover, because tensor networks have known limitations in their abilities to reproduce bulk geometry \cite{Bao_2015,Bao_2019}, it is all the more worthwhile to find a more general and explicit construction of a class of spacetime metrics capable of describing a Python's Lunch.

While we will arrive at postprandial pythons in the end, let us step back to the beginning and study the preprandial geometries that we initially advertised.
In \Sec{sec:AdS_dust_geometries}, we will motivate and derive the dust geometries we will consider in the remainder of the paper, in particular understanding its metric and geodesic behaviors for the case of a step-function dust distribution. In \Sec{sec:wormhole_construction}, we will construct a wormhole by cutting and pasting together portions of these dust geometries. In \Sec{sec:holoinfo}, for our step-function dust wormhole, we will investigate the entanglement wedge phases and calculate the well-studied entanglement wedge cross section and mutual information \cite{Umemoto_2018,Nguyen_2018, dutta2019canonical, Bao2_2019}, as well as the outer entropy~\cite{Nomura_2018,Bousso_2019}.
Finally, using the intuition gained from the step-function ansatz, in \Sec{sec:python} we will explicitly engineer a Python's Lunch geometry using a different dust distribution. In particular, we will numerically compute the wormhole diameter along its length and the entanglement wedge cross section. 
We discuss future directions and conclude in \Sec{sec:conclusions}.

\section{Geometries for AdS plus dust} \label{sec:AdS_dust_geometries}

We begin by deriving a new asymptotically AdS solution to the Einstein equation in 2+1 dimensions that contains pressureless dust.
After writing down the metric in \Sec{sec:metric}, we find the spacelike boundary-anchored geodesics for a simple dust distribution in \Sec{sec:geodesics}.

\subsection{Metric}\label{sec:metric}
Consider the Einstein equation for a three-dimensional spacetime with negative cosmological constant plus dust,\footnote{Throughout, we work with mostly-plus metric signature, sign conventions $R_{\mu\nu} = R^\rho_{\;\;\mu\rho\nu}$ and $R^\mu_{\;\;\nu\rho\sigma} = \partial_\rho \Gamma^\mu_{\nu\sigma} - \partial_\sigma \Gamma^\mu_{\nu\rho} + \Gamma^\mu_{\rho\alpha}\Gamma^\alpha_{\nu\sigma}-\Gamma^\mu_{\sigma\alpha}\Gamma^\alpha_{\nu\rho}$, and, unless otherwise specified, set $8\pi G = 1$.}
\begin{equation}
R_{\mu\nu}-\frac{1}{2}Rg_{\mu\nu}=T_{\mu\nu}+\frac{1}{\alpha^{2}}g_{\mu\nu}, \label{eq:Ein}
\end{equation}
where the only nonzero component of $T_{\mu\nu}$ is $T_{tt} \equiv \rho$ and we have written the cosmological constant as $\Lambda=-1/\alpha^2$. When $\rho = 0$, one has the ${\rm AdS}_{3}$ solution, which we write in Poincar\'e patch coordinates as
\begin{equation}
{\rm d}s^{2}=\frac{\alpha^{2}}{z^{2}}(-{\rm d}t^{2}+{\rm d}z^{2}+{\rm d}x^{2}),\label{eq:AdS}
\end{equation}
where the boundary is located at $z=0$.

When the dust density $\rho$ is not zero, but instead $\rho = \rho(x,z)$, we find that the class of metrics
\begin{equation}
{\rm d}s^{2}=\alpha^{2}{\rm sech}^{2}t\left[-{\rm d}t^{2}+\frac{1}{z^{2}}\left(c^{2}{\rm d}z^{2}+e^{2f(x,z)}{\rm d}x^{2}\right)\right] \label{eq:metric}
\end{equation}
is a solution to the Einstein equation \eqref{eq:Ein}.
Here, the dust density satisfies the differential equation
\be 
\rho=1-\frac{1}{c^{2}}+\frac{z}{c^{2}}\left[f'-z(f')^{2}-zf''\right],\label{eq:rho}
\ee
where we use ${}^\prime$ to denote $\partial_z$. We believe that this represents a new, large class of exact solutions to the Einstein equation; it does not appear explicitly in, e.g., Ref.~\cite{Garcia-Diaz:2017cpv} or among the metrics considered in \Ref{Remmen2018} and refs. therein.

The Cotton-York tensor $C_{\mu\nu}=\nabla^\alpha (R^{\beta}_{\mu}- \frac{1}{4}R \delta^\beta_{\mu})\epsilon_{\alpha\beta\nu}$ has nonzero components $C_{tz}=C_{zt} = c^2 e^{-f(x,z)}(\partial_x \rho)\cosh t$ and $C_{tx}=C_{xt} = -e^{f(x,z)} (\partial_z \rho)\cosh t$.
Defining the unit timelike vector $t_\mu = \alpha \sech t\, \partial_t$, the traceless Ricci tensor $R_{\mu\nu}-\frac{1}{3} R g_{\mu\nu}$ can be written as $(\rho \cosh^2 t/3\alpha^2) (g_{\mu\nu} + 3 t_\mu t_\nu)$.
Thus, in the notation of Refs.~\cite{Garcia-Diaz:2017cpv,Chow:2009km}, these spacetimes have a Cotton tensor that is a special case of Petrov type ${\rm I}$, with one eigenvalue vanishing, and have a traceless Ricci tensor of Petrov type ${\rm D}_{\rm t}$.

We see that $\rho$ is $t$-independent and, on slices of constant $t$, this metric looks like a warped version of hyperbolic space, with warp factor $e^{2f(x,z)}/c^{2}$. 
It might be desirable to set $c^{2}=1$, so that on constant-$t$
slices the $z$ coordinate in Eq.~\eqref{eq:metric} can be identified
with the $z$-coordinate in the hyperbolic plane in Eq.~\eqref{eq:AdS}.
For any desired density profile of $\rho(x,z)$, one then solves the differential equation $\rho=z\left[f'-z(f')^{2}-zf''\right]$
for $f(x,z)$ to obtain the metric at constant $x$, then stitches together the solutions for arbitrary $x$.

The metric \eqref{eq:metric} is time-reflection symmetric about $t=0$, since ${\rm sech}\,t$ is even. 
Thus, taking the $t=0$ slice, we can apply the
Ryu-Takayanagi (RT) formula to compute entropies of boundary regions by finding minimal spacelike geodesics.
The overall time-dependent factor in \Eq{eq:metric}, which vanishes when $t\rightarrow \pm \infty$, is reminiscent of FRW cosmologies written in conformal time, and indeed one could choose coordinates in which \Eq{eq:metric} represents a dust distribution that expands from a big bang, comes to rest at $t=0$, and recollapses to a big crunch.

It will be well motivated to take $c^{2}=1$, so that $f=0$ and $\rho = 0$ coincide. Defining $u = e^{f}$, for each value of $x$ we have the homogeneous, linear, second-order ordinary differential equation:
\begin{equation}
z^{2}u''-zu'+\rho(x,z)u=0.
\end{equation}
Defining $y=z^{2}$, we can recast this as
\begin{equation}
\frac{{\rm d}^2 u}{{\rm d}y^2} +\frac{\rho(x,\sqrt{y})}{4y^{2}}u=0. \label{eq:Schrodinger}
\end{equation}
Note that this is precisely the time-independent one-dimensional Schr\"odinger equation,
\be 
-\frac{{\hbar}^2}{2M}\frac{{\rm d}^2 \psi}{{\rm d}y^2} + V(y)\psi = E\psi,
\ee 
with the identifications
\be 
\begin{aligned}
u &\longleftrightarrow \psi \\
\rho &\longleftrightarrow \frac{8My^2}{\hbar^2}\left[E-V(y)\right].
\end{aligned}\label{eq:dictionary}
\ee 
This allows us to generate a large number of nontrivial dust solutions by directly translating solutions to the Schr\"odinger equation using the dictionary in \Eq{eq:dictionary}.
Interestingly, the energy condition $\rho > 0$ corresponds with the requirement that $E>V$, i.e., a ``classical''-like, oscillatory solution, viewing \Eq{eq:dictionary} in a WKB context.

Let us solve Eq.~\eqref{eq:Schrodinger} for the example density profile
\be \label{eq:stepprofile}
\rho=m\,\theta(z-\bar{z}),
\ee
i.e., empty AdS at small $z$ and constant dust density $m$ at $z>\bar{z}$. The comoving density of the dust is $\hat \rho_m = -T_t^{\;\;t} = (m/\alpha^2)\cosh^2 t$; we require the dust to be of positive mass density, so $m\geq 0$. 
Similarly, the comoving density and pressure associated with $\Lambda$ are $\hat \rho_\Lambda = -\hat p_\Lambda = -1/\alpha^2$.
The total comoving density is $\hat \rho = \hat\rho_m + \hat \rho_\Lambda$, which equals $(m-1)/\alpha^2$ within the $t=0$ surface. The spacelike volume of this surface, for $z>\bar z$, is infinite, so we must have $m<1$; were $m>1$, we would have divergent positive energy within the slice, which would overclose a $(2+1)$-dimensional spacetime~\cite{Deser:1983tn}.\footnote{We could define an effective equation-of-state parameter $w=\hat p/\hat \rho = (m \cosh^2 t - 1)^{-1}$. However, the requirement that $m\leq 1$ means that $w\leq -1$ within the $t=0$ surface.}
Defining
\be 
b \equiv \sqrt{1-m},
\ee
the solution is
\begin{equation}
f(z)=\theta(z-\bar z)(1-b)\log(z/\bar{z}),\label{eq:fheaviside}
\end{equation}
so the metric within the dusty region becomes
\begin{equation}
{\rm d}s^{2}=\alpha^{2}{\rm sech}^{2}t\left\{ -{\rm d}t^{2}+\frac{1}{z^{2}}\left[{\rm d}z^{2}+\left(\frac{z}{\bar{z}}\right)^{2(1-b)}{\rm d}x^{2}\right]\right\} .
\end{equation}
This metric has the nice feature that not only is $T_{tt}$ constant
for $z>\bar{z}$, but both $T_{\;\;t}^{t}$ and $R$ are $x$- and
$z$- independent for $z>\bar{z}$ (but not $t$-independent, with both
increasing exponentially at large $t$). At the maximal density $m=1$,
we have, for $z>\bar{z}$, a flat metric times a $t$-dependent conformal
factor:
\begin{equation}
{\rm d}s^{2}=\alpha^{2}{\rm sech}^{2}t\left(-{\rm d}t^{2}+{\rm d}Z^{2}+{\rm d}X^{2}\right).
\end{equation}
where $Z=\log z$ and $X=x/\bar{z}$.

Let us consider how our metric reduces to AdS. 
Without loss of generality, let us take $c>0$. Requiring $\rho = 0$ in \Eq{eq:rho}, we take $f(x,z)$ to be independent of $x$, which implies a general solution for $f(x,z)$:
\begin{equation}
f(z)=\log(c_+ z^{1+c}+c_{-}z^{1-c}),
\end{equation}
where $c_\pm$ are arbitrary constants.
The metric is thus:
\be 
{\rm d}s^{2}=\alpha^{2}{\rm sech}^{2}t\left\{ -{\rm d}t^{2}+\frac{1}{z^{2}}\left[c^{2}{\rm d}z^{2}+\left(c_+ z^{1+c}+c_{-}z^{1-c}\right)^{2}{\rm d}x^{2}\right]\right\} .\label{eq:metricc}
\ee 
Defining $\hat z = z/(c_+ z^{1+c} + c_- z^{1-c})$, we have 
\be 
\frac{{\rm d}\hat z^2}{\hat z^2} = \frac{c^2 {\rm d}z^2}{z^2}\left( \frac{c_+ z^c - c_- z^{-c}}{c_+ z^c + c_- z^{-c}}\right)^2.
\ee 
In order to avoid a coordinate singularity along the $\hat z$-direction, either $c_+$ or $c_-$ must vanish, so the metric reduces to
\be 
{\rm d}s^2 = \alpha^2 {\rm sech}^2 t \left[-{\rm d}t^2 + \frac{1}{\hat z^2} ({\rm d}\hat z^2+ {\rm d}x^2) \right].
\ee 
Defining $z'=\hat z\cosh t$ and $t'=\hat z\sinh t$, the metric takes the AdS
form:
\begin{equation}
{\rm d}s^{2}=\frac{\alpha^{2}}{z'^{2}}\left(-{\rm d}t'^{2}+{\rm d}z'^{2}+{\rm d}x^{2}\right).
\end{equation}

\begin{figure}[ht]
    \centering
 \includegraphics[width=\textwidth]{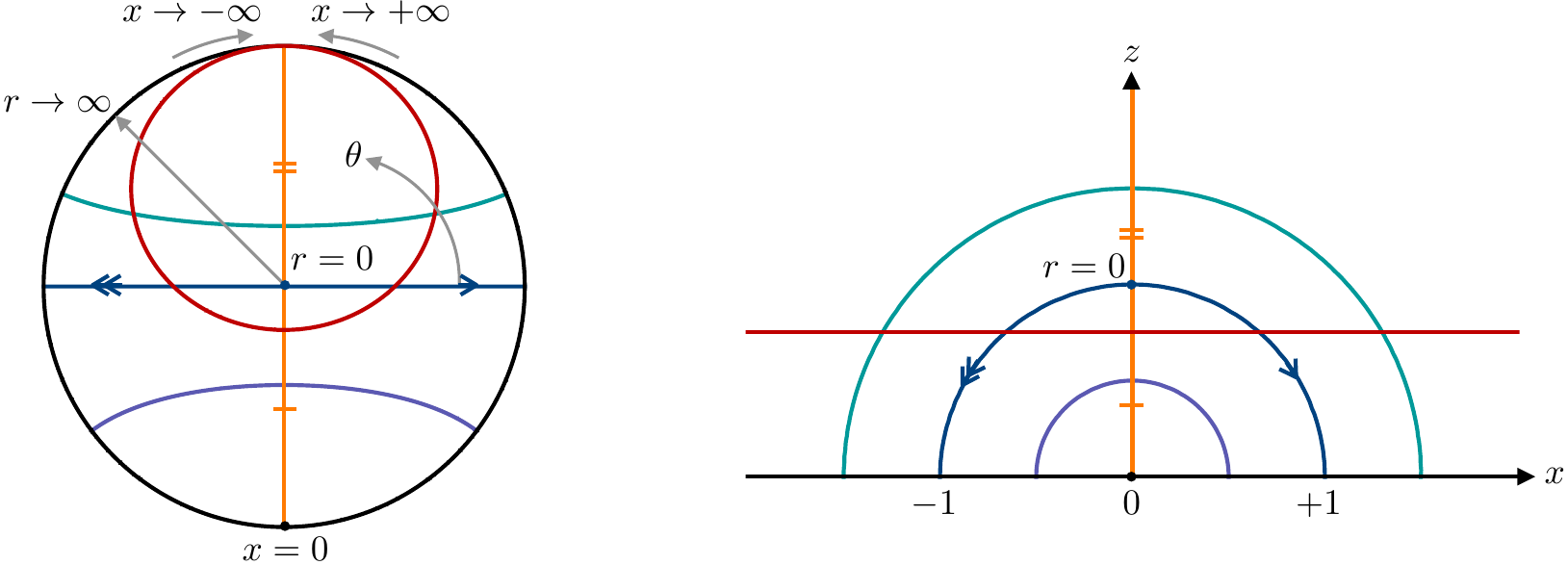}
    \caption{Various curves drawn in both the global disk (left) and the Poincar\'e patch (right).}
    \label{fig:patch_global}
\end{figure}

Note that a surface of constant $z$ in ${\rm AdS}_3$ is {\it not} a circle in the global coordinates $(\tau,r,\theta)$ of the hyperbolic disk, which we define implicitly via
\be 
(t,\;z,\;x) = \frac{1}{\cosh r \cos \tau - \sinh r \sin \theta}\times (\cosh r \sin \tau,\;1,\;\sinh r \cos \theta),
\ee
in which the ${\rm AdS}_3$ metric is
\be 
    {\rm d} s^2 = \alpha^2 \left( -\cosh^2 r ~ {
    \rm d}\tau^2 + {\rm d} r^2 + \sinh^2 r ~ {\rm d}\theta^2 \right).
\ee
For the purposes of drawing conformal diagrams, it is also useful to define $\cos \sigma = 1/\cosh r$, so that AdS\textsubscript{3} is a cylinder of radius $\pi/2$, and constant-$\tau$ surfaces are horizontal slices of the cylinder.
In these compact global coordinates, the metric reads ${\rm d}s^2 = \alpha^2 \sec^2 \sigma \left( -{\rm d} \tau^2 + {\rm d} \sigma^2 + \sin^2 \sigma ~ {\rm d}\theta^2 \right)$, and a surface of constant $z=\bar z$ within the $\tau = 0$ slice satisfies $\sec\sigma - \sin\theta\tan\sigma = 1/\bar z$.
See \Fig{fig:patch_global}, in which the dust for $\rho$ in \Eq{eq:stepprofile} will lie within (i.e., above) the red contour.

\subsection{Geodesics} \label{sec:geodesics}

An essential piece of data about the family of metrics we found in \Sec{sec:metric} are the geodesics, which we will need when constructing RT surfaces, wormholes, etc.
Let us take the metric in \Eq{eq:metric} with $c^2 = 1$, as before writing $u(x,z)=e^{f(x,z)}$ and ${}^\prime = \partial_z$, and consider a trajectory $p(\lambda) = (x(\lambda),z(\lambda))$, writing $\dot{}$ for differentiation with respect to the affine parameter $\lambda$. The geodesic equations within a constant-$t$ surface,
\begin{equation}
\begin{aligned}
\ddot{z}+\Gamma_{zz}^{z}\dot{z}^{2}+2\Gamma_{zx}^{z}\dot{z}\dot{x}  +\Gamma_{xx}^{z}\dot{x}^{2} &=0\\
\ddot{x}+\Gamma_{xx}^{x}\dot{x}^{2}+2\Gamma_{zx}^{x}\dot{z}\dot{x}  +\Gamma_{zz}^{x}\dot{z}^{2}&=0,
\end{aligned}
\end{equation}
then become
\begin{equation}
\begin{aligned}\ddot{z}-\frac{1}{z}\dot{z}^{2}+\left(\frac{u^2}{z}-u u'\right)\dot{x}^{2} & =0\\
\ddot{x}-2\left(\frac{1}{z}-\frac{u'}{u}\right)\dot{z}\dot{x} + \frac{\partial_x u}{u}\dot x^2 & =0.
\end{aligned}
\end{equation}

\begin{figure}[t]
    \centering
 \includegraphics[height=0.5\textwidth]{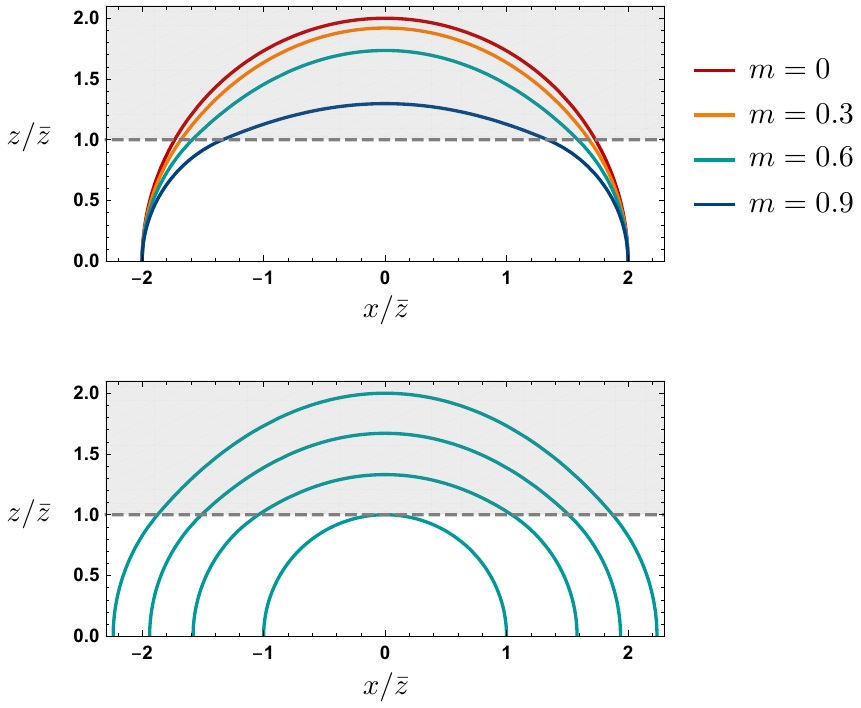}
    \caption{Illustration of the geodesics whose trajectories are given in \Eq{eq:geos}, for a step function dust density $m \,\theta(z-\bar z)$ (gray shading). For $z<\bar z$, the geodesics are the semicircles in $(x,z)$ coordinates given in \Eq{eq:geodesicsvac}. Geodesics reaching into the dusty region are increasingly pushed away from the dust for higher and higher dust densities.}
    \label{fig:geos}
\end{figure}

Let us consider the particular case we explored in \Sec{sec:metric}, where the dust density is $\rho = m\,\theta(z-\bar z)$ as in \Fig{fig:geos}, where the corresponding $f$ is given in \Eq{eq:fheaviside}.
In the region where $z<\bar{z}$, the geodesic equations reduce to those of empty AdS, and so the geodesics are arcs
of circles in $(x,z)$ coordinates: the boundary anchored geodesic centered at $x=x_0$ and extending to a depth $z=z_0$ at $\lambda = \lambda_0$ in the bulk satisfies
\be 
\begin{aligned}
x_{\rm vac}(\lambda) &= x_0 + z_0 \tanh [\sqrt{E_0}(\lambda-\lambda_0)] \\
z_{\rm vac}(\lambda) &= z_0 \sech [\sqrt{E_0}(\lambda-\lambda_0)],
\end{aligned}\label{eq:geodesicsvac}
\ee
where $E_0 = (\dot x^2 + \dot z^2)/z^2$ is a constant of motion; see \Fig{fig:geos}.
(The quantity $p_x = \dot x / z^2$ is another constant of motion, and we note the relation $z_0 = \sqrt{E_0}/p_x$ for later use.)
Regulating the geodesic by cutting off its ends at coordinate $z=\epsilon$, the proper length at constant $z$ subtended by the geodesic is $\simeq 2\alpha x_{0}/\epsilon$.
The $x$-coordinate subtended by the geodesic (i.e., distance in $x$ at $z=0$) is $\Delta x = 2z_0$. 
The geodesic reaches these endpoints at $\lambda = \lambda_0 \pm \lambda_\infty$, where $\lambda_\infty = E_0^{-1/2} {\rm arccosh}(z_0/\epsilon)$. The proper depth the geodesic reaches into the bulk is  $\alpha\int_{\epsilon}^{z_{0}}{\rm d}z/z=\alpha\log(z_{0}/\epsilon)$.
The length of the geodesic is
\be 
\begin{aligned}
\int_{\lambda_0-\lambda_\infty}^{\lambda_0 +\lambda_\infty}{\rm d}\lambda \sqrt{g_{\mu\nu}\dot x^\mu \dot x^\nu} &= 2\alpha \sqrt{E_0} \lambda_\infty 
\\& = 2\alpha\, {\rm arccosh}(z_0/\epsilon) = 2\alpha\,{\rm arccosh}(\Delta x/2\epsilon) \\
& =2\alpha\,\log(\Delta x/\epsilon) + {\cal O}((\epsilon/\Delta x)^2).
\end{aligned}\label{eq:Length}
\ee

However, if the geodesic hits
the dust boundary at $z=\bar{z}$, then the length in $x$ spanned by a geodesic will no longer equal twice its $z$-coordinate depth reached into the bulk, since the geodesic will deviate from the circular form for $z>\bar{z}$.
In the region where $z>\bar{z}$, the geodesic equations become:
\begin{equation}
\begin{aligned}\ddot{z}-\frac{\dot{z}^{2}}{z}+ b \left(\frac{z}{\bar{z}}\right)^{2(1-b)}\frac{\dot{x}^{2}}{z} & =0\\
\ddot{x}-\frac{2b}{z}\dot{z}\dot{x} & =0.
\end{aligned}
\end{equation}
The second geodesic equation is simply the conservation of momentum along the $x$-direction:
\begin{equation}
\dot{p}_{x}=0,
\end{equation}
where here, in the presence of dust,
\begin{equation}
p_{x}=\frac{\dot{x}}{z^{2}}\left(\frac{z}{\bar{z}}\right)^{2(1-b)}.\label{eq:pxdust}
\end{equation}
The first geodesic equation is then the single-variable ordinary differential
equation:
\begin{equation}
\frac{\ddot{z}}{z^{2}}-\frac{\dot{z}^{2}}{z^{3}}+b \left(\frac{\bar{z}}{z}\right)^{2(1-b)}zp_{x}^{2}=0.
\end{equation}
That is, defining $E=\dot{z}^{2}/z^{2}$ and noting that
${\rm d}E/{\rm d}z=\dot{E}/\dot{z}$ since we are now working with
a single variable, we have
\begin{equation}
\frac{{\rm d}E}{{\rm d}z}=-\frac{{\rm d}}{{\rm d}z}V(z),
\end{equation}
where now the effective potential is
\begin{equation}
V(z)=\bar{z}^{2}\left(\frac{z}{\bar{z}}\right)^{2b}p_{x}^{2}.
\end{equation}
The solution for $E$ is thus:
\begin{equation}
E=E_{0}-\bar{z}^{2}\left(\frac{z}{\bar{z}}\right)^{2b}p_{x}^{2},
\end{equation}
where $E_{0}$ is a constant.
Making a convenient definition of a rescaled $z$-coordinate,
\be 
\zeta(\lambda) = [z(\lambda)/\bar z]^b,
\ee
we have derived an energy conservation equation,
\begin{equation}
E_{0}=E+V= \left(\frac{\dot \zeta}{b \zeta}\right)^2 + \left(p_x \bar{z} \zeta\right)^2 ={\rm constant}.\label{eq:Econsdust}
\end{equation}
If we let $\zeta= \zeta_0$ denote the apex of the geodesic at which $\dot \zeta = 0$ and where we set $\lambda=0$, then $E_0 = (p_x \bar z \zeta_0)^2$
and the solution for which $\zeta(0) = \zeta_0$ is
\begin{equation} \label{eq:dustzgeod}
    \zeta(\lambda)  = \zeta_0 \sech (b\sqrt{E_0}\lambda).
\end{equation}
Then, from the definition of $p_x$, it follows that $\dot x = p_x \bar{z}^2 \zeta^2$,
which upon integration gives
\begin{equation} \label{eq:dustxgeod}
    x(\lambda) = x_0 + \frac{\bar{z} \zeta_0}{b} \tanh (b\sqrt{E_0}\lambda).
\end{equation}
The geodesic is an ellipse in $(x,\zeta)$:
\begin{equation}
    \zeta_0^2 = \left[\frac{b(x(\lambda)-x_0)}{\bar{z}} \right]^2 + \left[\zeta(\lambda) \right]^2.
\end{equation}
In these coordinates, the edge of the dust occurs at $\zeta = 1$.
The affine parameter $\pm \bar{\lambda}$ at which the geodesic impacts this surface is
\begin{equation} \label{eq:coshzeta0}
    \bar{\lambda} = \frac{1}{b\sqrt{E_0}}\arccosh \zeta_0 .
\end{equation}
From the elliptical relation, we further have
\begin{equation}
    x(\pm \bar{\lambda}) - x_0 = \pm \frac{\bar{z}}{b} \sqrt{\zeta_0^2 - 1} .
\end{equation}

Let us construct a full boundary anchored geodesic, centered at $x_0 = 0$.
We must match the vacuum geodesics in \Eq{eq:geodesicsvac}
at the dust interface with the dusty geodesics \eqref{eq:dustzgeod} and \eqref{eq:dustxgeod} for the same values of $E_0$ and $p_x$ in both solutions.
Recall that in the vacuum solution $z_0 = \sqrt{E_0}/p_x$, and from the dust solution we also have that $\sqrt{E_0}/p_x = \bar{z} \zeta_0$. Therefore, we must solve
\be 
\begin{aligned}
z_\mrm{vac}(\pm \bar{\lambda}) &= \bar z \\
x_\mrm{vac}(\pm \bar{\lambda}) &= \pm \frac{\bar{z}}{b} \sqrt{\zeta_0^2 - 1} \label{eq:step1}
\end{aligned}
\ee 
for $\lambda_0$ and $x_0$ appearing in \Eq{eq:geodesicsvac}.
The positive and negative roots give the right and left geodesic segments, respectively.
Since $E_0$ is the same throughout the spacetime, we can without loss of generality set it to unity via a constant rescaling of the affine parameter, which we will do henceforth.

Matching $z$ on the right side at $+ \bar{\lambda}$, \Eq{eq:step1} gives $\zeta_0 = \cosh(\bar{\lambda} - \lambda_0)$.
Comparing with \Eq{eq:dustzgeod}, we thus have $\cosh(\bar{\lambda}-\lambda_0) = \cosh(b  \bar{\lambda}) $,
which has two solutions:
\begin{equation} \label{eq:rightlambda0solution}
    \lambda_0 = (1 \pm b) \bar{\lambda} \qquad \text{(right vacuum segment)}.
\end{equation}
Similarly, matching at $- \bar{\lambda}$ gives $\cosh(-\bar{\lambda}-\lambda_0) = \cosh(- b  \bar{\lambda})$,
which has two solutions,
\begin{equation}
    \lambda_0 = -(1 \pm b) \bar{\lambda} \qquad \text{(left vacuum segment)}.
\end{equation}

Matching $x$ on the right segment at $+\bar{\lambda}$, \Eq{eq:step1} gives
\begin{equation}
    \begin{aligned}
    \frac{\bar{z}}{b} \sqrt{\zeta_0^2 - 1} = x_\mrm{vac}(+\bar{\lambda}) &= x_0 + \bar{z} \zeta_0 \tanh( \bar{\lambda} - \lambda_0) \\
    &= x_0 + \bar{z} \zeta_0 \tanh(\mp b \bar{\lambda}).
    \end{aligned}
\end{equation}
Since $\zeta_0 = \cosh(b \bar{\lambda})$, it follows that $\zeta_0 \tanh( b\bar{\lambda}) = \sqrt{\zeta_0^2 - 1}$,
so we arrive at
\begin{equation}
    x_0 = \left(\frac{1}{b} \pm 1\right) \bar{z} \sqrt{\zeta_0^2 - 1} \qquad \text{(right vacuum segment)},
\end{equation}
where the $\pm$ is in correspondence with \Eq{eq:rightlambda0solution}.
By inspection, we must take the negative solution to match the sign of the first derivative ($\dot \zeta(\bar{\lambda}) < 0$).
An analogous calculation for the left segment gives
\begin{equation}
    x_0 = - \left(\frac{1}{b} \pm 1\right) \bar{z} \sqrt{\zeta_0^2 - 1} \qquad \text{(left vacuum segment)},
\end{equation}
where again the correct solution is the negative root.

Putting this all together, the complete boundary-anchored geodesic is
\begin{equation}
\begin{aligned}
    x(\lambda) &= \left\{
    \begin{array}{ll}
    \displaystyle \pm x_0 + \bar{z} \zeta_0 \tanh (\lambda \mp \lambda_0 )  & \lambda \gtrless \pm \bar{\lambda} \\
    \displaystyle \frac{\bar{z} \zeta_0}{b} \tanh (b  \lambda) & |\lambda| \leq \bar{\lambda}
    \end{array} \right. \\
    z(\lambda) &= \left\{
    \begin{array}{ll}
    \displaystyle \bar{z} \zeta_0 \sech (\lambda \mp \lambda_0 )   & \lambda \gtrless \pm \bar{\lambda} \\
    \displaystyle \bar{z} \left[ \zeta_0 \sech (b  \lambda) \right]^{1/b} & |\lambda| \leq \bar{\lambda}
    \end{array} \right.,
\end{aligned}\label{eq:geos}
\end{equation}
where we have defined
\be 
\begin{aligned}
    \lambda_0 &= (1-b) \bar{\lambda} \\
    x_0 &= \left(\frac{1}{b} - 1\right) \bar{z} \sqrt{\zeta_0^2 - 1},
\end{aligned}\label{eq:lambda0x0}
\ee 
and for convenience we remind ourselves that
\begin{equation}
    \bar{\lambda} = \frac{1}{b} \arccosh \zeta_0.
\end{equation}
The bulk reach of the geodesic in $z$ is $\bar z \zeta_0^{1/b}$. The free parameters $(\bar{z},b,\zeta_0)$ control where the dust starts, the density of the dust, and how deep into the bulk the geodesic reaches; see \Fig{fig:geos} for an illustration.

It will be convenient to define $a = b^{-1} - 1$. The geodesic intersects the boundary ($z=0$) at $x = \pm \bar z  \left(\zeta_0 + a \sqrt{\zeta_0^{2} -1 }\right)$, thus subtending an $x$-coordinate of 
\be 
\Delta x = 2\bar z \left(\zeta_0 + a \sqrt{\zeta_0^2 - 1 }\right).\label{eq:deltaxdust}
\ee
Regulating the boundary at $z=\epsilon$, the geodesic intersects this surface at affine parameter $\lambda = \pm \lambda_\infty$, where
\be \label{eq:dustlambdainf}
\lambda_\infty =  \lambda_0 + {\rm arccosh}\left(\frac{\bar z \zeta_0}{\epsilon}\right).
\ee 
The length of the geodesic is 
\be 
\begin{aligned}
L &=\int_{-\lambda_\infty}^{\lambda_\infty}{\rm d}\lambda \sqrt{g_{\mu\nu}\dot x^\mu \dot x^\nu}  \\
&= 2\alpha \lambda_\infty \\
&= 2\alpha \left[ a\,\arccosh \zeta_0 +  \arccosh\left( \frac{\bar z \zeta_0}{\epsilon} \right)\right].
\end{aligned}
\ee
Note that we can rewrite $\zeta_0$ in terms of $\Delta x$ as 
\be 
\zeta_0 = \frac{-\Delta x + a\sqrt{\Delta x^2 + 4\bar z (a^2 - 1)}}{2\bar z(a^2 - 1)}.
\ee
If we hold $m$ fixed and send $\epsilon \rightarrow 0$, we clearly recover the logarithmic scaling of pure AdS.
However, if we hold $\epsilon$ (as well as $\Delta x$) fixed, and consider $m\rightarrow 1$, we obtain (assuming $\Delta x/\bar z > 2$, so that the geodesic actually reaches the dust section):
\be 
\begin{aligned}
L/\alpha &\rightarrow \left(\frac{\Delta x}{\bar z} - 2\right) + 2 \,{\rm arccosh}(\bar z/\epsilon ) \\
&\qquad + \frac{1-m}{24}\left(\frac{\Delta x}{\bar z} - 2\right) \left[ 8 - 2 \frac{\Delta x}{\bar z} - \frac{\Delta x^2}{\bar z^2} + \frac{6\bar z}{\sqrt{\bar z^2 -\epsilon^2}}\left(\frac{\Delta x}{\bar z}-2 \right) \right] + \cdots,
\end{aligned}\label{eq:Llinear}
\ee
where $\cdots$ indicates terms of higher order in $1-m$. We find linear scaling for maximal dust $m\simeq 1$, since in this limit the metric approaches Minkowksi in the dusty region, and the boundary anchored geodesic will track the $z= \bar z$ surface.

Thus, by choosing the value of $m$, we can tune the scaling of $L$ and thus the entanglement entropy of the boundary region subtended by the geodesic.
While $m=0$ corresponds to the usual logarithmic scaling of the entropy, the extremal $m\rightarrow 1$ limit gives us volume-law scaling of the entropy of the boundary region.
This gives us another way of understanding the pathology associated with setting $m>1$: it would lead to superextensive scaling of the entanglement entropy. Because the entanglement entropy is upper bounded by the logarithm of the dimension of the Hilbert space, something which is itself extensive (e.g., directly proportional to the number of lattice sites in a discretization), this would not be allowed for any quantum mechanical theory with a notion of space.

\section{Wormhole construction} \label{sec:wormhole_construction}

We would now like to construct a wormhole from our dusty asymptotically AdS$_3$ geometry.
In a spatial slice of empty AdS, one can construct a two-sided wormhole by taking a quotient of the space by a hyperbolic isometry \cite{Aminneborg:1997pz,Brill:1998pr}.
This essentially amounts to cutting the hyperbolic plane along two boundary-anchored geodesics and then gluing along the cuts to form the wormhole.
For more general spacetimes, this gluing procedure for codimension-two surfaces $\gamma_1$ and $\gamma_2$ is consistent and results in a smooth wormhole geometry provided that junction conditions are met \cite{EngelhardtWall}, specifically, that $\gamma_1$ and $\gamma_2$ are isometric and that there exist null normals to $\gamma_{1,2}$ with vanishing expansion and twist.
We will choose $\gamma_{1,2}$ to be two semicircular, spacelike, boundary-anchored geodesics within the empty AdS portion of the geometry, centered about $\pm x_{\rm bdy}$ and with depth $z_{\rm bdy}$.
Since these geodesics are extremal surfaces and reflections of each other across the $z$-axis, the two geodesics satisfy the spacelike analogues of the conditions of \Ref{EngelhardtWall} and may thus be glued together to construct a wormhole.

\begin{figure}[t]
    \centering
   \includegraphics[width=0.7\textwidth]{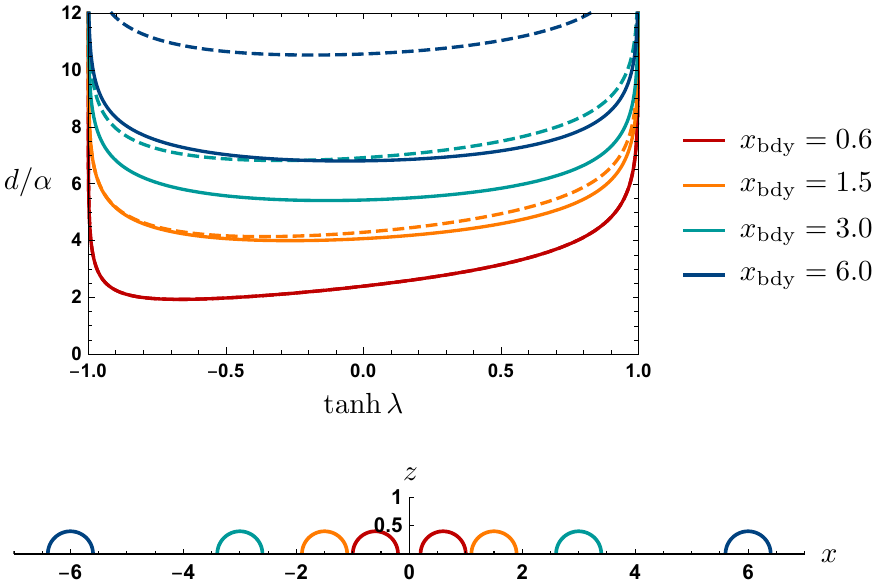}
    \caption{Wormhole construction described in text, where we identify pairs of geodesics within the dust-free portion of the spacetime, for several choices of spacing $x_{\rm bdy}$. In this example, we choose the dust to begin at $\bar z = 1$ and let the geodesics reach a depth of $z_{\rm bdy} = 0.5$, so that each subtends an $x$-coordinate distance of $1$ in the boundary. The wormhole width function $d(\lambda)$ is plotted in the case of a spacetime with $m=0.9$ dust for $z>1$ (dotted lines) and for pure AdS (solid lines). We see that the dust enlarges the girth of the wormhole, as expected.}
    \label{fig:geoid}
\end{figure}

\begin{figure}[t]
    \centering
   \includegraphics[width=0.4\textwidth]{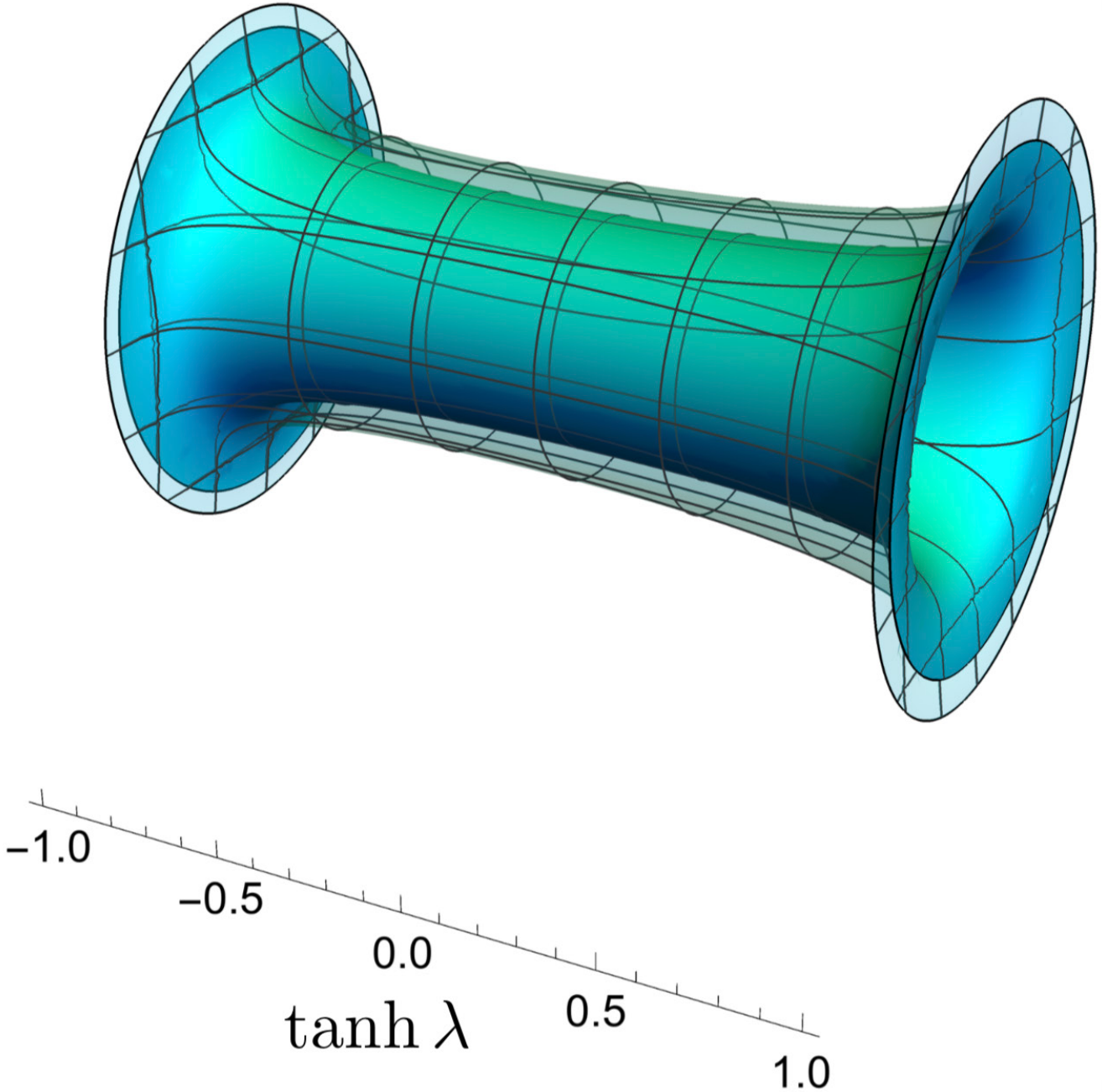}
    \caption{Wormhole obtained from the construction in \Fig{fig:geoid} in the case $x_{\rm bdy} = 3$. The vacuum wormhole is the inner solid surface, and the wormhole with dust is the outer translucent surface.}
    \label{fig:wormhole}
\end{figure}

Let $\lambda=- \infty$ (respectively, $+\infty$) correspond, for both geodesics, to the endpoint nearest (respectively, furthest) from $x=0$.
The two geodesics' points at fixed $\lambda$ are some $(\pm x, z)$ in Poincar\'e patch coordinates.
These two points can themselves be connected by a geodesic centered on $x=0$, described by \Eq{eq:geos}, for some value of $z_0 = \bar{z} \zeta_0^{1/b}$ for which the two points are intersected for some $\lambda$ in \Eq{eq:geos}; call these values $\hat z_0$ and $\hat \lambda$, respectively.
Finding $\hat z_0$ (or equivalently, $\hat \zeta_0=(\hat z_0/\bar z)^b$) and $\hat \lambda$ amounts to inverting \Eq{eq:geos}. We can then compute the distance between $(\pm x,z)$ and thus the proper width $d(\lambda)$ of the wormhole at each affine parameter $\lambda$, which will be simply $2\alpha \hat \lambda$; see Figs.~\ref{fig:geoid} and \ref{fig:wormhole}.\footnote{Defining the proper length differential ${\rm d}\tau = {\rm d}\lambda \sqrt{g_{\mu\nu}\dot x^\mu \dot x^\nu}$, the expression in the square root defines the ``tick rate'' of the affine parameter $\lambda$ relative to proper time. On an affinely parameterized geodesic, this rate is constant and related to the energy of the system: $2\alpha E_0 = {\rm d} \tau/{\rm d}\lambda = \sqrt{g_{\mu\nu} \dot x^\mu \dot x^\nu}$.  Were we to rescale $\lambda$ by a constant, we would by definition have to rescale $\sqrt{E_0}$ by the inverse constant, which leaves the length invariant. In the case in text, we have chosen units such that $E_0 = 1$.} 

Let us take $x>0$ without loss of generality. For $z>\bar z$, we have:
\be 
\begin{aligned}
\hat \zeta_0 &= \sqrt{\zeta^2 + \left(\frac{b x}{\bar z} \right)^2} \\
\hat \lambda &= \frac{1}{b} \arcsinh(bx/\zeta \bar z) .
\end{aligned}
\ee
For $z\leq \bar z$ but $x^2 + z^2 \geq \bar z^2$, so that the apex of the geodesic still sits within the dust region, we find that $\hat x_0 \equiv x_0(\hat \zeta_0)$, given in \Eq{eq:lambda0x0}, must satisfy a quadratic equation:
\begin{equation}\label{eq:nicequadratic}
    (1-a^2) \hat x_0^2 + 2a^2 x \hat x_0 - a^2(x^2 + z^2 - \bar{z}^2) = 0.
\end{equation}
Solving and selecting the correct branch, we find:
\be 
\begin{aligned}
\hat \zeta_0 &= \sqrt{1 + \left[ \frac{- a x+\sqrt{x^2 + (z^2 - \bar{z}^2)(1-a^2)}}{\bar{z} (1- a^2)}  \right]^2}\\
\hat \lambda &=\arccosh\left(\frac{\bar z \hat\zeta_0}{z} \right)+ a\,\arccosh \hat\zeta_0. 
\label{eq:inverse}
\end{aligned}
\ee

Finally, in the case where $x^2 + z^2 < \bar{z}^2$, the geodesic connecting $(\pm x,z)$ only probes the empty AdS region and we have the vacuum solution:
\be 
\begin{aligned}
   \hat z_0 &= \sqrt{x^2+z^2} \\
   \hat \lambda &= \arcsinh (x/z) .
\end{aligned}
\ee 

Since the wormhole is constructed by cutting and gluing along $\gamma_1$ and $\gamma_2$, the geodesic distance between pairs of identified points (i.e., points at the same affine parameter value on both geodesics) is a measure of the width of the wormhole.
A plot of this width as we go from the outermost points on $\gamma_{1,2}$ to the innermost points is shown in \Fig{fig:geoid} for several choices of $\gamma_{1,2}$.
We also show the width function for a dust-free space-time, and consequently see that the presence of dust results in a uniformly wider wormhole.
We can also visualize the wormhole by plotting a surface of revolution of this geodesic width function, as shown in \Fig{fig:wormhole}, although we note that this should simply be thought of a visual aid, since it is not an embedding diagram.

\section{Holographic information}\label{sec:holoinfo}

\begin{figure}[t]
    \centering
  \includegraphics[width=0.7\textwidth]{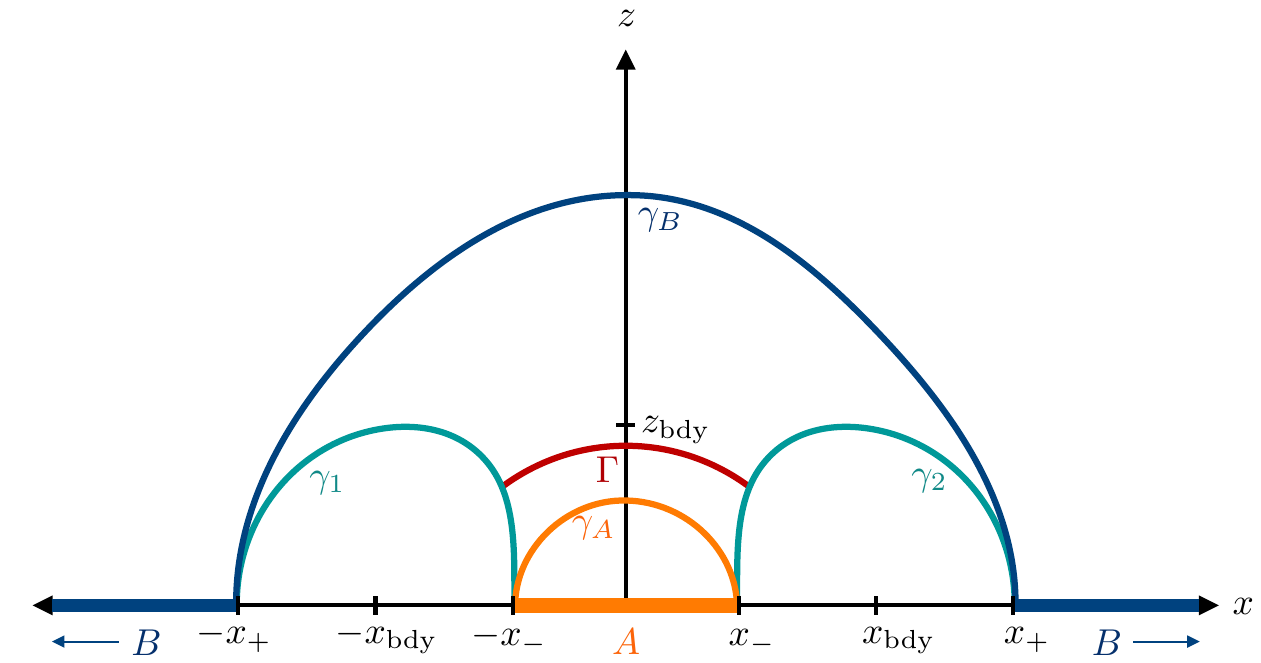}
    \caption{Summary of the labels defined in Secs.~\ref{sec:wormhole_construction} and \ref{sec:holoinfo}. When $|\gamma_1|+|\gamma_2|<|\gamma_A|+|\gamma_B|$, the entanglement wedge for $AB$ is in the connected phase, and the entanglement wedge cross section $\Gamma$ connects $\gamma_1$ with $\gamma_2$.}
    \label{fig:labels}
\end{figure}

We now turn to the computation of holographic quantities for the AdS plus dust spacetime that we have been investigating (the metric \eqref{eq:metric} with the dust profile \eqref{eq:stepprofile}).
Let $x_+ > x_- > 0$, and on the $t=0$ slice, define two boundary subregions, $A$ and $B$, as follows:
\begin{equation}
    A = (-x_-, x_-), \qquad B = (-\infty, -x_+) \cup (x_+,\infty).
\end{equation}
Let $\gamma_A$ (respectively, $\gamma_B$) be the boundary-anchored geodesic that subtends $A$ (respectively, $B$).
If we choose $(x_+ - x_-)/2 < \bar{z}$, then the geodesics that subtend the intervals $(-x_+, -x_-)$ and $(x_-,x_+)$ are the geodesics $\gamma_1$ and $\gamma_2$ that we used to contruct a wormhole in \Sec{sec:wormhole_construction}, with $x_{\rm bdy} = (x_+ + x_-)/2$ and $z_{\rm bdy} = (x_+ - x_-)/2$; see \Fig{fig:labels}.

Since the $t=0$ slice is a moment of time-reflection symmetry, $\gamma_A$ and $\gamma_B$ are the extremal HRT surfaces of $A$ and $B$.
If there exists a holographic CFT state in two dimensions that is dual to our dusty spacetime, then the entropies of the reduced states on $A$ and $B$ are given by
\begin{equation}
    S(A) = \frac{|\gamma_A|}{4 G}, \qquad S(B) = \frac{|\gamma_B|}{4G\hbar}.
\end{equation}
The HRT surface of the joint system $AB$ will be either $\gamma_A \cup \gamma_B$ or $\gamma_1 \cup \gamma_2$, whichever has smaller length.
In the first case, the entanglement wedge of $AB$ is just the union of the entanglement wedges of $A$ and $B$ separately.
In the latter case, the entanglement wedge of $AB$ is a single connected component, and it coincides with the part of the spacetime that we wrapped up to construct a wormhole in \Sec{sec:wormhole_construction}.

We will primarily be interested in the connected phase, and so in \Sec{subsec:connected}, we establish the conditions under which the entanglement wedge of $AB$ is connected.
We then go on to investigate various holographic properties of the connected entanglement wedge, in particular focusing on the entanglement wedge cross section in \Secs{subsec:ewxsec}{subsec:ep-mi}.
We end with a calculation of outer entropy in \Sec{subsec:outerentropy}.

\subsection{Conditions for a connected entanglement wedge} \label{subsec:connected}

The condition for a connected entanglement wedge is $|\gamma_A| + |\gamma_B| > |\gamma_1| + |\gamma_2|$.
Let us continue to suppose that $z_{\rm bdy} = (x_+ - x_-)/2 < \bar{z}$, so that $\gamma_1$ and $\gamma_2$ lie entirely in the dust-free region.
Then with a $z$-coordinate cutoff at $z = \epsilon$, it follows that (cf. \Eq{eq:Length})
\begin{equation}
    |\gamma_1| = |\gamma_2| = 2 \alpha \arccosh \left( \frac{z_{\rm bdy}}{\epsilon}\right) = 2 \alpha \log \left( \frac{2z_{\rm bdy}}{\epsilon}\right) + O(\epsilon^2) .
\end{equation}

Next, consider $\gamma_A$ (all results for $\gamma_B$ follow by making the substitution $x_- \rightarrow x_+$).
If $x_- < \bar{z}$, then $\gamma_A$ also lies entirely in the dust-free region, and so
\begin{equation}
    |\gamma_A|_{x_- < \bar{z}} = 2 \alpha \arccosh \left( \frac{x_-}{\epsilon}\right) = 2 \alpha \log \left( \frac{2x_{-}}{\epsilon}\right) + O(\epsilon^2) .
\end{equation}
If $x_- > \bar{z}$, then $|\gamma_A| = 2 \alpha \lambda_\infty$, where $\lambda_\infty$ is given by \Eq{eq:dustlambdainf}:
\begin{equation}
\begin{aligned}
    |\gamma_A|_{x > \bar{z}} &= 2 \alpha \left[ \arccosh\left( \frac{\bar{z}\hat{\zeta}_0(x_-)}{\epsilon} \right) + a \arccosh \hat{\zeta}_0(x_-) \right] \\
    &= 2 \alpha \left[ \log\left( \frac{2\bar{z}\hat{\zeta}_0(x_-)}{\epsilon} \right) + a \arccosh \hat{\zeta}_0(x_-) \right] + O(\epsilon^2).
\end{aligned}
\end{equation}
In the above, we use $\hat{\zeta}_0(x_-)$ to denote $\hat{\zeta}_0$, as defined in \Eq{eq:inverse}, evaluated at $z = \epsilon$ and $x \simeq x_-$.

There are three cases for where $\gamma_A$ and $\gamma_B$ lie.
For each of these cases, let us compute the quantity 
\begin{equation}
C = \frac{1}{2\alpha} \left(|\gamma_A| + |\gamma_B| - |\gamma_1| - |\gamma_2|\right).
\end{equation}
When it is positive, then the entanglement wedge of $AB$ is in the connected phase and $C$ is proportional to the mutual information $I(A:B)$.
Note that this categorization holds for all allowed values $0 \leq m \leq 1$, encompassing the strict area-law regime ($m=0$, no dust) up to the volume-law regime ($m=1$, critical density dust); the $m$-dependence enters through $\hat \zeta_0 (x_\pm)$.

\begin{enumerate}
    \item If $x_- < \bar{z}$ and $x_+ < \bar{z}$, then $\gamma_A$ and $\gamma_B$ both lie entirely in the dust-free region:
    \begin{equation} \label{eq:case1}
    \begin{aligned}
        C &= \log \left( \frac{2x_-}{\epsilon} \right) + \log \left( \frac{2x_+}{\epsilon} \right) - 2 \log \left( \frac{2z_{\rm bdy}}{\epsilon} \right) + O(\epsilon^2) \\
        &= \log\left( \frac{x_- x_+}{z_{\rm bdy}^2}\right) + O(\epsilon^2).
    \end{aligned}
    \end{equation}
    
    \item If $x_- < \bar{z}$ and $x_+ \geq \bar{z}$, then $\gamma_A$ is in the dust-free region and $\gamma_B$ crosses into the dust:
    \begin{equation} \label{eq:case2}
    \begin{aligned}
        C &= \log \left( \frac{2x_-}{\epsilon} \right) + \log \left( \frac{2\bar{z} \hat{\zeta}_0(x_+)}{\epsilon} \right) - 2 \log \left( \frac{2z_{\rm bdy}}{\epsilon} \right) + a \arccosh \hat{\zeta}_0(x_+) + O(\epsilon^2) \\
        &= \log\left( \frac{x_- \bar z \hat{\zeta}_0(x_+) }{z_{\rm bdy}^2} \right) + a \arccosh \hat{\zeta}_0(x_+) + O(\epsilon^2).
    \end{aligned}
    \end{equation}
    
    \item If $x_- \geq \bar{z}$ and $x_+ \geq \bar{z}$, then both $\gamma_A$ and $\gamma_B$ cross into the dust:
    \begin{equation} \label{eq:case3}
    \begin{aligned}
        C &= \log \left( \frac{\bar{z} \hat{\zeta}_0(x_-)}{\epsilon} \right) + \log \left( \frac{\bar{z} \hat{\zeta}_0(x_+)}{\epsilon} \right) - 2 \log \left( \frac{z_{\rm bdy}}{\epsilon} \right) \\& \;\;\; +  a \left(\arccosh \hat{\zeta}_0(x_-) + \arccosh \hat{\zeta}_0(x_+)\right) + O(\epsilon^2) \\
        &= \log\left(  \frac{\bar z^2 \hat{\zeta}_0(x_-) \hat{\zeta}_0(x_+)}{z_{\rm bdy}^2} \right) +  a \left(\arccosh \hat{\zeta}_0(x_-) + \arccosh \hat{\zeta}_0(x_+)\right) + O(\epsilon^2).
    \end{aligned}
    \end{equation}
\end{enumerate}

Since $z_{\rm bdy} < \bar{z}$ by assumption, it follows that taking $x_- > \bar{z}$ is sufficient to guarantee that the entanglement wedge is in the connected phase.
This is fairly clear in the first case (recall that $x_+ > x_-$).
For the other two cases, recall that $\bar z$ corresponds to the coordinate location $\zeta = 1$, and so the apexes satisfy $\hat{\zeta}_0(x_\pm) > 1$.

\subsection{Entanglement wedge cross section} \label{subsec:ewxsec}

The entanglement wedge cross section $\Gamma$ is the minimal surface that partitions the entanglement wedge of $A$ and $B$ into two regions, one adjacent to $A$ but not $B$ and one adjacent to $B$  but not $A$. It was first defined by Refs.~\cite{Umemoto_2018, Nguyen_2018}, and its area in Planck units, denoted by $E_W(A:B)$, was initially conjectured to be dual to the entanglement of purification, $E_P(A:B)$ \cite{Terhal2002TheEO}.
Several other candidates for its holographic dual have since been proposed, including reflected entropy \cite{dutta2019canonical,Bao5_2019}, odd entropy \cite{Tamaoka:2018ned}, and logarithmic negativity \cite{Kudler-Flam:2018qjo}; consensus about the correct holographic dual, however, has yet to emerge. In addition, $E_W(A:B)$ has been generalized to conditional and multipartite configurations in Refs.~\cite{Bao5_2019,Chu:2019etd,Bao3_2019, Bao4_2018,Umemoto2_2018}.

Understanding the holographic dual of $E_W$ is of interest to the gravity community because it would constitute an additional nontrivial entry in the dictionary between geometric and entropic quantities in AdS/CFT. Furthermore, since the proposed entropic duals of $E_W$ appear to coincide in the ground state, it is valuable to compute $E_W$ in examples away from the AdS vacuum (although we will not carry out any entropic calculations here, since we are not investigating an explicit CFT dual). From the perspective of the quantum information community, it is interesting that the fairly elaborate entropic quantities listed above are supposedly dual to a geometric quantity that is as simple as $E_W$.

Returning to our example, let us continue to suppose that $\gamma_{1,2}$ lie in the dust-free region.
Let us take $x_- > \bar{z}$, so that the entanglement wedge is in the connected phase and proceed to calculate $E_W(A:B)$.
The cross section $\Gamma$ is therefore a curve with one endpoint on $\gamma_1$ and the other on $\gamma_2$.

First, we note that the $\mathbb{Z}_2$ symmetry guarantees that $\Gamma$ must itself be $\mathbb{Z}_2$-symmetric. Suppose this were not the case. Then the $\mathbb{Z}_2$ symmetry guarantees that there are instead two surfaces $\Gamma$ and $\Gamma'$, where $\Gamma$ and $\Gamma'$ are each not $\mathbb{Z}_2$ symmetric, but for which $\Gamma\leftrightarrow \Gamma'$ under the $\mathbb{Z}_2$. Let $\Gamma\cap \gamma_1 = a_1$, $\Gamma \cap \gamma_2 = a_2$, $\Gamma' \cap \gamma_1 = b_1$, $\Gamma' \cap \gamma_2 = b_2$, and $\Gamma \cap \Gamma' =c$, so $\Gamma$ is the geodesic arc $a_1ca_2$, while $\Gamma'$ is the geodesic arc $b_1cb_2$.
By hypothesis, $|\Gamma| = |\Gamma'|$ is the minimal length for any surface connecting a point in $\gamma_1$ with one in $\gamma_2$.
Without loss of generality, suppose $|a_1c| \leq |b_1c|$. Then the surface $a_1cb_2$ satisfies $|a_1cb_2| \leq |\Gamma|$. (If instead $|b_1c|<|a_1c|$, then replace $a_1cb_2$ with $b_1ca_2$ here and in what follows.)
But since $\Gamma \neq \Gamma'$ by hypothesis, $a_1cb_2$ is not a geodesic due to the kink at $c$.
Hence, there must be another, geodesic, surface $\Gamma_0$ connecting $a_1$ and $b_2$ with $|\Gamma_0|$ {\it strictly} less than $|a_1cb_2|$.
This contradicts the hypothesis that $\Gamma$ and $\Gamma'$ are entanglement wedge cross sections; this contradiction completes the proof that $\Gamma$ must be $\mathbb{Z}_2$-symmetric.

Since $\Gamma$ is $\mathbb{Z}_2$-symmetric, it is one of the geodesics connecting $(\pm x(\lambda),z(\lambda))$ at fixed $\lambda$ on $\gamma_2$ and $\gamma_1$, respectively; these are precisely the surfaces we considered in \Sec{sec:wormhole_construction} to construct wormholes.
Since the length of this geodesic is $2 \alpha \hat \lambda(x(\lambda),z(\lambda))$, we can find $\Gamma$ by extremizing $\hat \lambda$ with respect to the affine parameter $\lambda$.

Beginning with \Eq{eq:inverse}, define $\beta \equiv \sqrt{\hat{\zeta}_0^2-1}$ and write
\begin{equation} \label{eq:d2}
    \hat \lambda(\lambda) = \arccosh \left( \frac{\bar{z}}{z} \sqrt{\beta^2+1}  \right) + a \arccosh \left( \sqrt{\beta^2+1} \right).
\end{equation}
Note that $z = z(\lambda)$ is a function of $\lambda$, and $\beta$ itself is a function of $x(\lambda)$ and $z(\lambda)$.
Expanding out ${\rm d}\hat{\lambda}/{\rm d}\lambda = 0$, we may thus write
\begin{equation}
    0 = \left(\frac{\dee x}{\dee \lambda}\frac{\partial \beta}{\partial x} + \frac{\dee z}{\dee \lambda} \frac{\partial \beta}{\partial z} \right) \frac{\partial \hat \lambda}{\partial \beta} + \frac{\dee z}{\dee \lambda} \frac{\partial \hat{\lambda}}{\partial z} .
\end{equation}
Working through the derivatives and making use of \Eq{eq:nicequadratic} (note that $\hat x_0 = a \bar{z} \beta$, cf. \Eq{eq:lambda0x0}), we eventually find that
\begin{equation} \label{eq:wedgestep1}
    \frac{{\rm d} x/{\rm d} \lambda}{{\rm d} z/{\rm d} \lambda} = \frac{x - \hat x_0}{z}
\end{equation}
However, $x(\lambda)$ and $z(\lambda)$ were chosen so that they lie entirely in the dust-free region,
\begin{equation} \label{eq:wedgevac}
    \begin{aligned}
       x(\lambda) &= x_{\rm bdy} + z_{\rm bdy} \tanh \lambda \\
z(\lambda) &= z_{\rm bdy} \sech \lambda,
    \end{aligned}
\end{equation}
and so
\begin{equation} \label{eq:wedgestep2}
    \frac{{\rm d} x/{\rm d} \lambda}{{\rm d} z/{\rm d} \lambda} = -\frac{1}{\sinh \lambda} = - \frac{z}{x - x_{\rm bdy}} .
\end{equation}
Combining \Eqs{eq:wedgestep1}{eq:wedgestep2}, we thus find that
\begin{equation}
    (x-\hat{x}_0)(x-x_{\rm bdy}) + z^2 = 0.
\end{equation}
Making use of \Eq{eq:wedgevac}, we find that the solution, $\lambda_*$, satisfies the quadratic equation
\be \label{eq:lambdaprimequad}
(z_{\rm bdy} + x_{\rm bdy} \tanh \lambda_*)^2 + a^2 [(\bar{z}^2+z_{\rm bdy}^2)\tanh^2 \lambda_* -z_{\rm bdy}^2] = 0.
\ee 
Solving the quadratic and choosing the right branch, we have 
\be 
\tanh \lambda_* = - \frac{z_{\rm bdy}}{x_{\rm bdy}}\left[\frac{1+a\sqrt{1+(a^2 - 1) \frac{\bar{z}^2+z_{\rm bdy}^2}{x_{\rm bdy}^2}}}{1+a^2 \frac{\bar{z}^2+z_{\rm bdy}^2}{x_{\rm bdy}^2}}\right].\label{eq:tanhlambda}
\ee
We then can evaluate $\hat{\lambda}(\lambda_*)$ using \Eq{eq:d2} to find the entanglement wedge cross section. We have:
\be 
|\Gamma|/2\alpha = {\rm arccosh}\left(-\frac{1}{\tanh\lambda_*}\right) + a \,{\rm arccosh} \left(-\frac{z_{\rm bdy}/\bar{z}}{\sinh\lambda_*}\right),\label{eq:bigGamma}
\ee
where $\tanh \lambda_*$ is given in \Eq{eq:tanhlambda}.

\subsection{Relation to entanglement of purification and mutual information} \label{subsec:ep-mi}

One of the interesting inequalities obeyed by the entanglement wedge cross section $E_W(A:B)=|\Gamma|/4G\hbar$ that is also obeyed by the entanglement of purification $E_P(A:B)$ is that it is lower-bounded by half the mutual information, $I(A:B)/2$. In holographic systems, the difference between these two quantities appears to be nonperturbatively large in a number of examples~\cite{MSSZ}. This makes intuitive sense, because of the difference in operational definition between the mutual information and the entanglement of purification. While the entanglement of purification is the measure of the number of Bell pairs needed to prepare a state with local operations and asymptotically vanishing communication,\footnote{It is illuminating here to note the difference with the distillable entanglement, which first assumes local operations and arbitrarily large amounts of classical communication, and second is a statement about the number of Bell pairs that can be derived from a state, rather than the number needed to create a state. The restriction to vanishing amounts of communication is a quite constraining one that makes the preparation process inefficient.} the mutual information says nothing about this quantity, and is instead simply a total correlation measure between the subsystems $A$ and $B$. It is simple to show from this operational sense that the number of Bell pairs needed to prepare the state must be at least sufficient to prepare this amount of correlation between $A$ and $B$, and it is also easy to see how in general one could require many more Bell pairs to prepare the state in the constrained manner that characterizes $E_P$. The difference (or the ratio) between the two quantities is a parametrization of the efficiency of preparing holographic density matrices from Bell pairs, and the demonstrated large gap speaks to the relative lack of efficiency in these protocols.

In our dust example, we indeed continue to see this large difference between the entanglement wedge cross section's area and half the mutual information.
From our earlier calculations, we wish to compare $I(A:B)/2 = \alpha C / 4G\hbar$ to $E_W(A:B) = |\Gamma|/4G\hbar$, the width of the entanglement wedge cross section. As a tractable example, let us consider a dust that approaches extremality, $m\rightarrow 1$, and with $x_->\bar{z}$, so that the length of $\gamma_A$ and $\gamma_B$ are as given in \Eq{eq:Llinear}:
\be 
\begin{aligned}
|\gamma_A|/2\alpha &= \frac{x_-}{\bar{z}} - 1 + \arccosh (\bar{z}/\epsilon) \\
|\gamma_B|/2\alpha &= \frac{x_+}{\bar{z}} - 1 + \arccosh (\bar{z}/\epsilon),
\end{aligned}
\ee
so
\be
\begin{aligned}
C &= \frac{x_- + x_+}{\bar{z}} - 2 + 2\arccosh(\bar{z}/\epsilon) - 2\arccosh((x_+ - x_-)/2\epsilon) \\
&= 2\left[\frac{x_{\rm bdy}}{\bar{z}} - 1  - \log \left(\frac{z_{\rm bdy}}{\bar{z}}\right)\right] + O(\epsilon^2).
\end{aligned}
\ee
On the other hand, taking the $m\rightarrow 1$ limit in \Eq{eq:bigGamma}, we find that the entanglement wedge cross section is also linear in $x_{\rm bdy}$:
\be 
|\Gamma|/2\alpha = \frac{x_{\rm bdy}}{\bar{z}} - \sqrt{1+\frac{z_{\rm bdy}^2}{\bar{z}^2}} + {\rm arccsch}\left(\frac{z_{\rm bdy}}{\bar{z}}\right).
\ee
We thus find
\be 
4G\hbar \cdot \frac{1}{2\alpha}\left[E_W(A:B) - \frac{1}{2}I(A:B)\right] = 1-\sqrt{1+\frac{z_{\rm bdy}^2}{\bar{z}^2}} + {\rm arccsch}\left(\frac{z_{\rm bdy}}{\bar{z}}\right) + \log\left(\frac{z_{\rm bdy}}{\bar{z}}\right),
\ee
which is strictly positive for all $z_{\rm bdy}$ in the range allowed by our construction ($z_{\rm bdy}/\bar z < 1$).
Conversely, if we take $m=0$, we have
\be 
4G\hbar \cdot \frac{1}{2\alpha}\left[E_W(A:B) - \frac{1}{2}I(A:B)\right] = \arccosh\left(\frac{x_{\rm bdy}}{z_{\rm bdy}}\right) - \log\sqrt{\frac{x_{\rm bdy}^2}{z_{\rm bdy}^2}-1},\label{eq:m0}
\ee
which is also positive.

We already see that the difference between $E_W(A:B)$ and $I(A:B)/2$ is never infinitesimal in these particular cases.
Numerical analysis reveals that the tightest bound occurs when $m=1$, $x_{\rm bdy} = 2 \bar{z}$, and $z_{\rm bdy} = \bar{z}$, in which case
\begin{equation}
    \min ~ 4G\hbar \cdot \frac{1}{2\alpha} \left[E_W(A:B) - \frac{1}{2}I(A:B)\right] =  1-\sqrt{2} + \log(1+\sqrt{2}) \approx 0.467.
\end{equation}
The difference is unbounded in the $x_- x_+/\bar{z}^2 \leq 1$, $x_+/\bar{z}>1$ regime as well as the $x_+ / \bar{z} < 1$ regime, in the limit where $x_{\rm bdy} \rightarrow z_{\rm bdy}$.

\subsection{Outer entropy} \label{subsec:outerentropy}

It will be illuminating to connect this class of geometries to the outer entropy construction of Refs.~\cite{Nomura_2018,Bousso_2019}.
Given a compact, closed surface $X$ in some geometry, the outer entropy $S^{(\rm outer)}$ is defined as the area (divided by $4G\hbar$) of the largest black hole that can be constructed in the spacetime, subject to holding geometry fixed in the wedge outside of $X$ (defined with respect to null geodesics) and requiring that $T_{\mu\nu}$ satisfies the null energy condition, excluding the cosmological constant.
As such, outer entropy is a coarse-grained holographic entropy that quantifies uncertainty when only a portion of a holographic spacetime is known.
Furthermore, it has been shown in special cases \cite{EngelhardtWall} that outer entropy coincides with the area of holographic screens \cite{Bousso2015a,Bousso2015b}. These codimension-one hypersurfaces obey a monotonicity theorem that is analogous to Hawking's area theorem---that the area of all event horizons can only increase when the null energy condition is satisfied \cite{Hawking1972}.
However, they differ from event horizons in that they are quasi-locally defined and occur in generic spacetimes, even those that do not contain black holes.
An algorithm for computing the outer entropy in spherically-symmetric~\cite{Nomura_2018}---and subsequently for general~\cite{Bousso_2019}---spacetimes has been developed, which we apply here.

For the class of spacetimes we find in \Sec{sec:metric}, it is natural to investigate $S^{(\rm outer)}$ for $X$ a surface at constant $z=z_*$.
While this choice of $X$ is noncompact, we can adopt a natural generalization of the outer entropy to an outer entropy density ${\cal S}^{(\rm outer)}$ per unit area, in analogy with entropy density per unit area for a planar black hole.

We can define the ingoing and outgoing orthogonal null congruences for the metric in \Eq{eq:metric}, with tangent vectors
\be 
k^{(\pm)}_\mu = \frac{\alpha}{\sqrt{2}}{\rm sech}\, t\,(1,\pm c/z,0)
\ee 
and expansions
\be 
\theta_{\pm} = \nabla_\mu k^{(\pm)\mu} = \frac{1}{\alpha}\left[\sqrt{2}\sinh t \pm \frac{1}{\sqrt{2} c} \left(-1+z f'\right) \right].
\ee
The shear, intrinsic curvature, and twist of $X$ all vanish.

Following \Ref{Bousso_2019}, we then have the outer entropy density:
\be 
{\cal S}^{(\rm outer)} = \frac{1}{4G\hbar} \sqrt{1-\frac{2\theta_+ \theta_-}{\Lambda}}.
\ee 
Let us take $X$ to lie in the time-symmetric $t=0$ hypersurface of interest for applying the RT formula and further take it to lie in a region of constant dust density $m$ for $z>\bar z$, where $0\leq m \leq 1$ as before in order to simultaneously satisfy energy conditions and avoid overclosure of the geometry. We therefore take $f$ as in \Eq{eq:fheaviside}, in which case the expansions become, for $z_* > \bar z$,
\be 
\theta_{\pm} =\pm  \sqrt{\frac{1-m}{2\alpha^2}}.
\ee 
We then have $\lambda = 1/(1-m)$, so that the entropy density becomes
\be 
{\cal S}^{(\rm outer)} = \frac{\sqrt{m}}{4G\hbar}.
\ee
We have zero outer entropy when $m=0$, i.e., empty AdS, consistent with \Ref{Nomura_2018}.
In the limit of maximal dust density $m=1$, the outer entropy density equals that of a black hole.
This is in keeping with the behavior of boundary-anchored geodesics that we found in the step-function case in \Sec{sec:geodesics}; just as for AdS black holes, the boundary-anchored geodesics do not penetrate the dusty region when $m=1$.

\section{The Python's Lunch} \label{sec:python}

As we saw in \Sec{sec:geodesics}, geodesics in the class of dust geometries that we derived tend to be dust-phobic.
That is, positive-energy dust lengthens distances along the $x$-direction, so paths of minimal length try to avoid passing through dusty regions for long distances, preferring vacuum AdS.
The result, as illustrated in \Fig{fig:wormhole}, is that the presence of dust in the geometry enlarges the girth of the wormhole constructed via the procedure in \Sec{sec:wormhole_construction}.
However, we found that the uniform, step-function dust profile explored in Secs.~\ref{sec:AdS_dust_geometries}~and~\ref{sec:wormhole_construction} enlarges the {\it entire} wormhole: rather than a Python's Lunch metric exhibiting a bulge, this geometry corresponds to simply a larger width along the entire length of the wormhole.

Let us therefore instead posit a different dust profile, using the lessons learned from the simpler solution in order to engineer a true Python's Lunch geometry. 
As before, let $\gamma_1$ and $\gamma_2$ be two boundary-anchored geodesics that we glue together to construct a wormhole.
Again suppose that $\gamma_{1,2}$ are reflections of each other across the $z$-axis, so that the geodesic length between pairs of glued points is the circumference of the corresponding wormhole section.
Reconsider the case of pure AdS for a moment, and imagine running through the family of these geodesics, starting with the points on $\gamma_{1,2}$ nearest to the origin and ending with the points farthest from the origin.
The geodesic length of course decreases to a minimum value---the throat of the wormhole---before increasing again as the wormhole flares out.
Since dust increases geodesic length, if we were to instead put a {\it localized} region of dust somehwere in the region swept out by these geodesics, then it seems plausible that it could upset the monotonicity of the geodesics' lengths, thus producing a bulge in the wormhole.

\begin{figure}[t]
    \centering
 \includegraphics[height=0.5\textwidth]{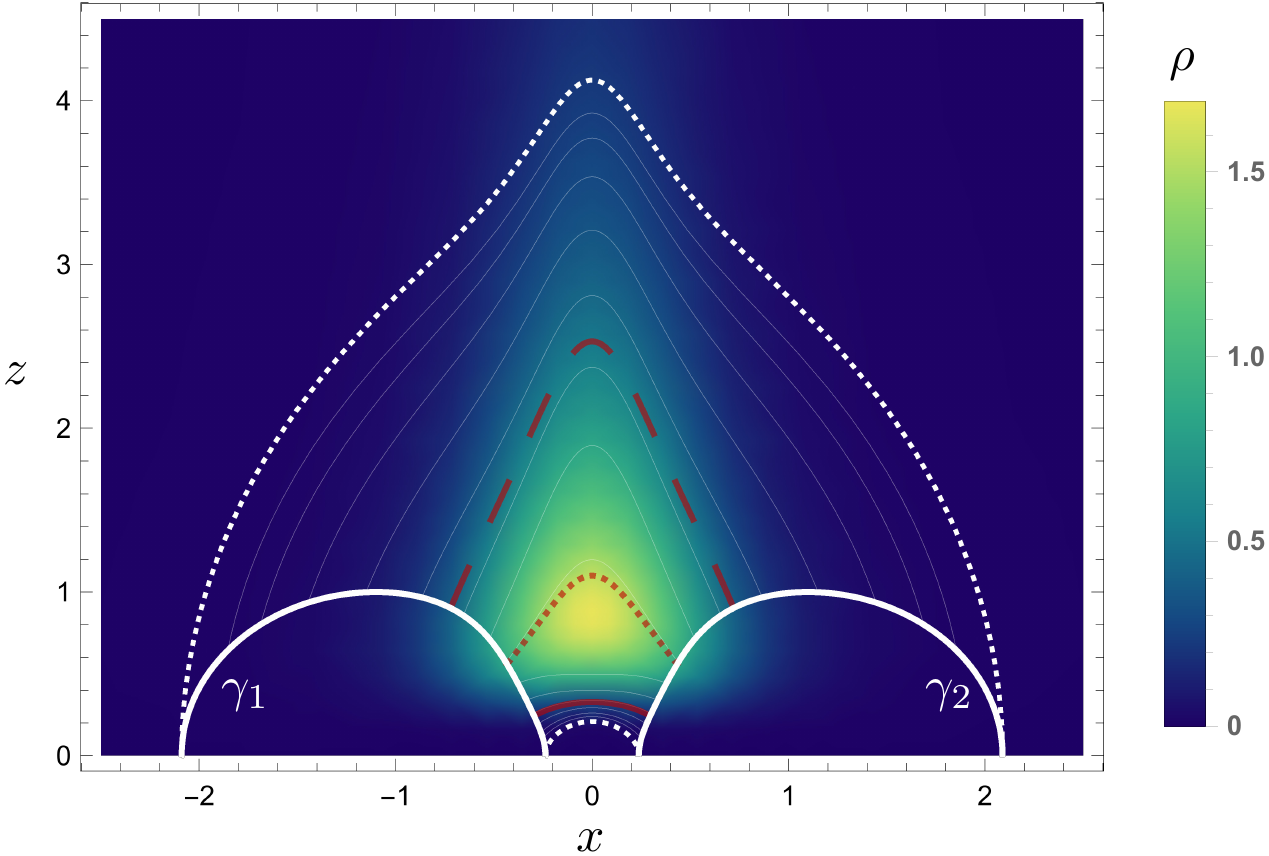}
    \caption{Construction of a Python's Lunch wormhole, using the metric in \Eq{eq:metric} with dust density $\rho$ given in \Eq{eq:dustprofile}. The boundary anchored geodesics $\gamma_{1,2}$ are identified (thick white lines). The wormhole region lies between the geodesics subtending regions $A$ and $B$ (dotted white lines). The circumference of the wormhole at a given affine parameter is found from the length of the geodesic connecting the analogous points on $\gamma_{1,2}$; examples are shown by the thin, translucent lines. This circumference exhibits three extrema (red lines): two minima (solid and dashed), one of which is the entanglement wedge cross section (solid), and one local maximum (dotted), which constitutes the bulge of the Python's Lunch.}
    \label{fig:constructionplot}
\end{figure}

This intuition indeed turns out to be correct.
In the general metric we exhibited in \Eq{eq:metric}, let us set $c^2$ and $\alpha$ to unity for simplicity, and define
\be \label{eq:simplef}
f(x,z)=\frac{8z^2}{(1+4z^2)(1+x^2)^4}.
\ee
Then the dust density profile is
\be 
\rho(x,z) = \frac{256 z^4 \left[x^2(2+x^2)(2+2x^2+x^4)(1+4z^2) + 4z^2 \right]}{(1+x^2)^8 (1+4z^2)^4},\label{eq:dustprofile}
\ee
which is everywhere nonnegative and describes a localized dust overdensity, as shown in \Fig{fig:constructionplot}. Since $f\rightarrow 0$ as $z\rightarrow 0$ or as $x\rightarrow \infty$, the metric approaches AdS at the boundary, as well as away from $x=0$. The $\mathbb{Z}_2$ symmetry enforced by $f(x,z)=f(-x,z)$ ensures that our earlier construction applies.
While one could in principle first specify $\rho(x,z)$ and then determine $f(x,z)$ numerically by solving the differential equation \eqref{eq:rho}, this is a convenient closed-form expression that produces a local blob of dust.

\begin{figure}[ht]
    \centering
  \includegraphics[width=0.4\textwidth]{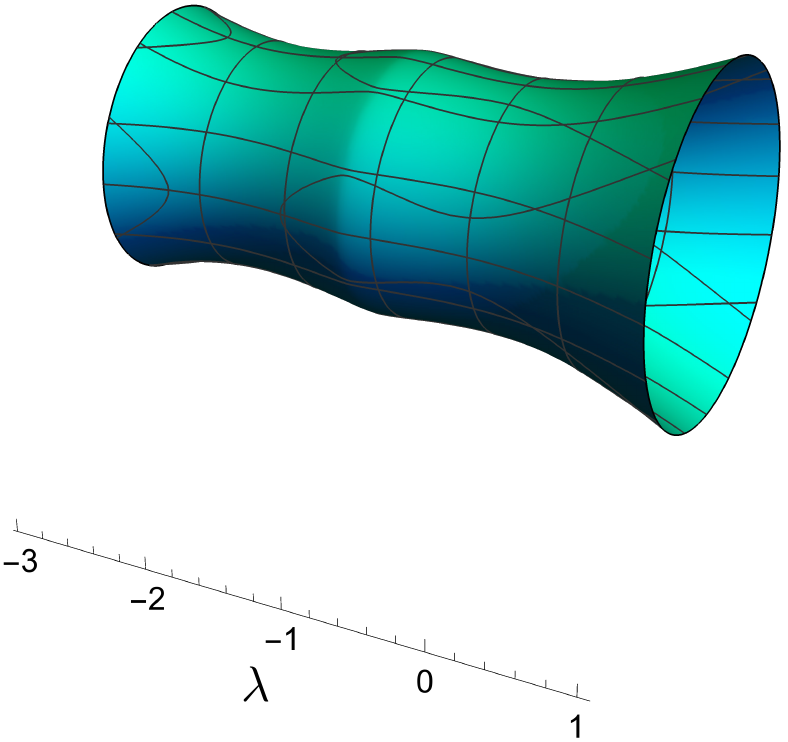}
    \caption{The Python's Lunch wormhole constructed from the AdS plus dust metric depicted in \Fig{fig:constructionplot}. Here as well, $\lambda$ is the affine parameter along the boundary geodesics $\gamma_{1,2}$.}
    \label{fig:tubefigure}
\end{figure}

While we know the metric exactly for this dust distribution, we will eschew computing geodesics analytically, since doing so would in this case require solving transcendental, rather than algebraic, equations.
However, we can numerically compute geodesic length and construct a wormhole. Let us take $x_0 = \pm 1.1$ and $z_0=1$ for the boundary-anchored geodesics $\gamma_{1,2}$, as depicted in \Fig{fig:constructionplot}. We find that by identifying these geodesics, we see that we have indeed engineered a bulge, and we obtain the Python's Lunch geometry shown in \Fig{fig:tubefigure}; as in \Fig{fig:wormhole}, \Fig{fig:tubefigure} is not an embedding diagram, but rather depicts the wormhole width function $d(\lambda)$ for this construction as a surface of revolution.
The ends of the wormhole flare out as $\gamma_{1,2}$ approach the boundary, just as in \Fig{fig:wormhole}.
However, near the middle of the wormhole, there is a local maximum and two local minima.\footnote{A Mathematica notebook that reproduces these numerics is provided as a supplemental file.}
One can check numerically that $|\gamma_1| + |\gamma_2| < |\gamma_A| + |\gamma_B|$, so our entanglement wedge is indeed in the connected phase, with boundary $\gamma_1 \cup \gamma_2$, and that $4G\hbar\, E_W \approx 3.3$, while $4G\hbar\,I(A:B) \approx 0.08$, so $E_W > I(A:B)/2$ as required.
The various order-one coefficients appearing in \Eq{eq:simplef} affect properties of the Python's Lunch such as the size of the bulge in relation to the outer minimal surfaces or the location of the bulge, but our ansatz does not offer us detailed control over these quantities.
In particular, one must be careful that the resulting energy density $\rho$ is everywhere positive.
Rather, this ansatz is a simple proof of principle for constructing the Python's Lunch, where the tools (i.e., the large class of $(2+1)$-dimensional metrics) to do so in more generality were developed in \Sec{sec:AdS_dust_geometries}.

\section{Conclusions} \label{sec:conclusions}

The class of metrics constructed in this paper definitively demonstrates that the Python's Lunch exists in well-defined spacetimes with positive-density matter. It is indeed possible for the minimal surfaces nearest each of the two asymptotic boundaries of a wormhole to not be the same surface, but rather enclose a larger space between them.

More broadly, our class of metrics that we investigated for the step-function dust density exhibits tunable scaling of the entanglement of a boundary subregion, from logarithmic (when $m=0$) to volume-law (when $m=1$).
The broad customizability of this class of geometries, allowing for an arbitrary $\rho(x,z)$, provides a multitude of fruitful future directions of study for constructions of interest to holography, including custom-shaped wormholes, deformation of RT surfaces, and behavior reminiscent of black holes.

A potential future direction with respect to the Python's Lunch geometry is to determine whether the bulge region can be made arbitrarily large. In Ref.~\cite{Brown:2019rox}, geometrical properties of the Python's Lunch geometry were argued to provide the difference between restricted and unrestricted complexity as argued for by Susskind et al. in Ref.~\cite{Brown_2016} as opposed to that argued by Harlow and Hayden in Ref.~\cite{Harlow_2013}. In particular, the maximal cross section of the bulge was in some sense the ``intermediate expression swell'' of the quantum circuit, while the volume and length approximated the total number of gates and total depth of the circuit, respectively, needed to perform specific operations. It would be interesting if our method of constructing dust geometries---or more general methods---could constrain any of these three parameters, as doing so could bound the degree to which arguments concerning the exponential separation between the two notions of complexity can be encapsulated in a relativistic setting. 
Another possibility would be to adapt the methods of, e.g., \Ref{BCFK_2019} to investigate the phases of bulk reconstructability induced by these bumpy wormhole geometries.
We leave these and other avenues to future work.

\begin{center} 
{\bf Acknowledgments}
\end{center}
\noindent 
We thank Raphael Bousso, Thomas Hertog, and Don Marolf for useful discussions and comments. 
N.B. is supported by the National Science Foundation under grant number 82248-13067-44-PHPXH, by the Department of Energy under grant number DE-SC0019380, and by the Computational Science Initiative at Brookhaven National Laboratory.
A.C.D. is a Postdoctoral Fellow (Fundamental Research) of the Research Foundation -- Flanders (Fonds Wetenschappelijk Onderzoek), File Number 12ZL920N.
G.N.R. is supported by the Miller Institute for 
Basic Research in Science at the University of California, Berkeley.

\bibliographystyle{utphys-modified}
\bibliography{refs.bib}

\end{document}